\documentstyle[12pt]{article}

\begin{document}

\renewcommand\baselinestretch {1.01}
\large\normalsize

\begin{center}

\large
{\bf {Transverse Dynamics of Spin-$\frac{1}{2}$ $XX$ Chain\\ 
      with Correlated Lorentzian Disorder}}

\vspace{3mm}       

\normalsize
{\bf {\underline{O. Derzhko} and T. Krokhmalskii}}\\
\normalsize
Institute for Condensed Matter Physics,
L'viv-11, 290011, Ukraine

\end{center}

\vspace{5mm}

\normalsize
\noindent
Statistical mechanics of spin models with disorder is an important field
of condensed matter theory onto which much attention is focused at present
time. All experimentally accessible magnetic systems with localized
magnetic moments are always affected by randomness and the understanding
of disorder effects within the frames of simple models can be of much use
in analysis of experimental data.
Usually two types of disorder are considered,
namely, the diagonal and off-diagonal disorder (see for example [1]).
However, it is evident that for alloys 
as well as for topologically disordered systems the random diagonal
and off-diagonal matrix elements involved into the Hamiltonian of the system
are connected and one faces the correlated diagonal and off-diagonal 
disorder. Starting from 1974 there exists a model of tight-binding 
electrons with correlated Lorentzian disorder the density of states for 
which can be found exactly [2]. Latter on the method elaborated in Ref. [2] 
was used for obtaining the density of magnon states for 
the spin-$\frac{1}{2}$ $XX$ chain with correlated Lorentzian disorder [3]. 
However, the obtained analytic results pertain only to thermodynamics and an 
interesting question about spin correlations and their dynamics still 
remains open. The aim of the present paper is to discuss this problem using 
a numerical approach developed recently [4,5]. 

We consider a chain of $N$ spins $1\over2$ governed by the Hamiltonian
$$
H=\sum_{n=1}^N\Omega_ns_n^z
+\sum_{n=1}^{N-1}J_n(s_n^xs_{n+1}^x+s_n^ys_{n+1}^y).
$$
The exchange couplings are assumed to be random independent variables each 
with the Lorentzian probability distribution
$$
p(J_n)=\frac{1}{\pi}
\frac{\Gamma}{(J_n-J_0)^2+\Gamma^2},
$$
where $J_0$ is the mean value of exchange coupling and $\Gamma$ controls the 
strength of disorder. The transverse fields are assumed to be related to the 
exchange couplings in the following way
$$
\Omega_n-\Omega_0=a\left(\frac{J_{n-1}-J_0}{2}+\frac{J_n-J_0}{2}\right),
$$
where $\Omega_0$ is the mean value of transverse field at site. We are 
interested in the $zz$ time-dependent correlation functions
$\langle s_j^z(t)s_{j+n}^z\rangle$ and the random-averaged 
transverse dynamic structure factor and susceptibility defined by the 
formulae
$$
\overline{S_{zz}(\kappa,\omega)}
=\sum_{n=1}^N
{\mbox{e}}^{{\mbox{i}}\kappa n}
\int_{-\infty}^{\infty}{\mbox{d}}t
{\mbox{e}}^{-\epsilon\mid t\mid}
{\mbox{e}}^{{\mbox{i}}\omega t}
\left[
\overline{\langle s_j^z(t)s_{j+n}^z\rangle}
-\overline{\langle s_j^z\rangle\langle s_{j+n}^z\rangle}
\right],\;\;\;\epsilon\rightarrow +0,
$$
$$
\overline{\chi_{zz}(\kappa,\omega)}
=\sum_{n=1}^N
{\mbox{e}}^{{\mbox{i}}\kappa n}
\int_0^{\infty}{\mbox{d}}t
{\mbox{e}}^{{\mbox{i}}(\omega+{\mbox{i}}\epsilon)t}
\frac{1}{{\mbox{i}}}
\overline{\langle\left[s_j^z(t),s_{j+n}^z\right]\rangle},
\;\;\;\epsilon\rightarrow +0,
$$
respectively.

An idea of numerical computations of these quantities is 
explained in details in Refs. [4,5].
We considered chains of
$N=300$ spins with $J_0=-1,$ $\Omega_0=0.5$ and $\Gamma=0.1$
at the low temperature $\beta=1000.$
We computed correlation functions
$\langle s_{150}^z(t)s_{150+n}^z\rangle
-\langle s_{150}^z\rangle\langle s_{150+n}^z\rangle$
with $n=0,\pm 1,\ldots,\pm 100$ for the times up to $t=200,$
put $\epsilon=0.01$ and averaged the $zz$ dynamic structure factor 
and susceptibility over 
3000 random realizations.
To understand the accuracy of the obtained results we performed 
additional calculations revealing the effects of finite $N$ and $n$ 
for the considered times $t$ at the taken value of temperature $\beta$. We 
also compared the numerical results with the analytic ones obtained for 
non-random case [6].

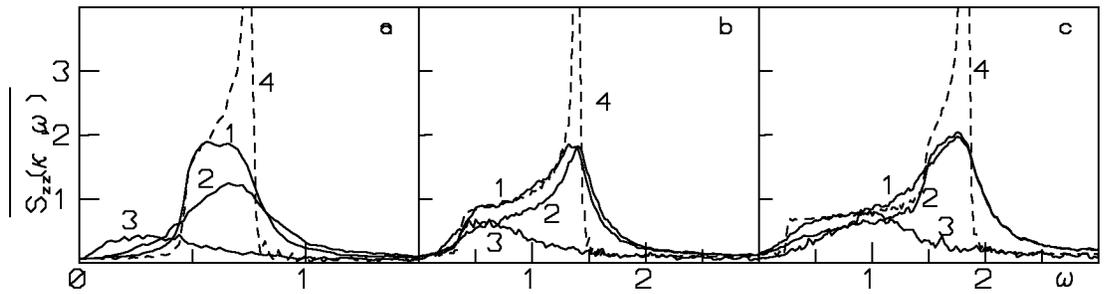
\begin{figure}[h] \setlength{\unitlength}{1.in}
\begin{picture}(7.5,1.5)(0,0)
\put(0.1,1.5){\special{em:graph ber_1.pcx}}
\end{picture}
\caption[]{
 $\overline{S_{zz}(\kappa,\omega)}$ vs. $\omega$ for several values 
of $\kappa$.}
\end{figure} 

In Fig. 1 we plotted the frequency dependence of the random-averaged dynamic 
structure factor for several values of wave vector $\kappa=\frac{\pi}{4}$ 
(a), $\kappa=\frac{\pi}{2}$ (b), $\kappa=\frac{3\pi}{4}$ (c). One can easily 
see the differences between the cases $a=-1.01$ (1), $a=1.01$ (2), and the 
case when the transverse field at site is random Lorentzian variable 
independent on the values of exchange couplings at neighbouring sites (3). 
The frequency profiles for non-random case $\Gamma=0$ are depicted by dashed 
lines (4). The well-pronounced differences in the plotted frequency shapes 
can be explained bearing in mind that $zz$ dynamics is conditioned by 
excitation of two magnons and taking into consideration the changes in the 
density of magnon states caused by the correlated disorder with opposite 
signs of $a$ and the not correlated disorder.

\vspace{5mm}

\noindent
[1] J. M. Ziman,
Models of Disorder (Cambridge University Press, 1979).

\noindent
[2] W. John and J. Schreiber,
Phys. Status Solidi B  {\bf 66}, 193 (1974).

\noindent
[3] O. Derzhko and J. Richter,
Phys. Rev. B {\bf 55}, 14298-14310 (1997).

\noindent
[4] O. Derzhko and T. Krokhmalskii,
Ferroelectrics {\bf 192}, 21-27 (1997).

\noindent
[5] O. Derzhko and T. Krokhmalskii,
Phys. Rev. B {\bf 56}, 11659-11665 (1997).

\noindent
[6] S. Katsura, T. Horiguchi, and M. Suzuki,
Physica {\bf 46}, 67-86 (1970).

\end{document}